\documentclass[12pt]{article}

\usepackage[letterpaper, margin=1.15in]{geometry}
\usepackage{natbib}
\usepackage{amsmath}
\usepackage{upgreek}
\usepackage{comment}
\usepackage{pdflscape}
\usepackage{afterpage}
\usepackage{caption,subcaption}
\usepackage{color,soul}
\usepackage{graphicx}

\title{``Keep the fight unfair": Military rhetoric in quantum technology}
\author{Emma McKay\footnote{emma.m.mckay@mcgill.ca}}

\begin{document}
\maketitle


\begin{abstract}
	Doing quantum ethics properly will require detailed socio-political analysis of the technologies and the organizations trying to build them. In this paper, I contribute to this task by analysing the public rhetoric of American military stakeholders in the quantum industry. I look at Air Force Research Laboratory involvement in the 2020 Quantum 2 Business conference, where they were the main sponsor. A critical thematic analysis shows a focus on enacting the violence of war, maintaining narratives that the Air Force provides a secure future for Americans, and marrying quantum technology with the aesthetics of war. I contextualize this with anti-imperialist theory, arguing that this rhetoric and the desire for quantum arms aligns with the reproduction of existing violent power structures. Insights about this example of military involvement in quantum spaces should help orient nascent critical quantum ethics interventions.
\end{abstract}

\section{Introduction}


	Technology is often developed for warfare, like atomic and hydrogen bombs, radar, transistors, and drones \citep{Schuster2015,VanKeuren1997,Douglas2018}. Modern warfare kills, injures, and displaces millions around the globe annually with enormous environmental cost \citep{Belcher2020,Benjamin2018a,Mohanty2008}. Many of these wars are started by nations with imperial interests, like the United States of America \citep{Immerwahr2019,Christiansen2020,Harvey2003}. To consider the ethics of new technologies, the question of whether and how they may be used to support militarized empire is critical.

	How are imperial militaries engaged with quantum science and technology (QST)? Should QST support any part of imperial warfare? Engaging with these questions is crucial to address the main question of ethics of emerging technology: should we build it or not?

	I engage with these questions in this paper by investigating the stated socio-technical aims and the implied values of the United States' Air Force Research Laboratory (AFRL) in QST. I draw on anti-imperial and linguistic analysis of American military bodies to ground a critical thematic analysis of publicly available quantum-related materials from the AFRL. I root this in a critical socio-political approach to ethics, where ethics happens by disrupting power structures, rather than by simply not making them worse.

	I come to this work as a scholar of science and technology studies (a social science discipline) with a master's in physics. I use literature and tools from sociology and related disciplines. I am trained, as is typical in the social sciences, to pay attention to positionality. Using the first person in writing is part of my attempt to make this positionality clear to the reader. For these reasons, this paper will not read like a typical quantum technology paper. 


	The case examined here is just one example of military involvement in QST; there is interest from the US Army \citep{Kott2018,Smith2020a}, the US Naval Research Laboratory \citep{OfficeOfNavalResearch2021}, the Australian Army \citep{AustralianArmy2021}, and the Canadian Armed Forces \citep{NationalDefenceCanada2021}, among others. The discussion here offers a point of reflection for those participating in QST: should militaries be involved with QST? It also offers a warning: they already are.

\section{Personal motivations of the author}
	\label{sec:author}


	In the feminist social sciences, scholars recognize that all research comes from people and is shaped by their particular social situation and motivation \citep{Haraway1988}. Providing this information explicitly adds useful context much in the same way that literature reviews do. So, here I discuss why I am attending to militarism in QST.

	I studied quantum information in my undergraduate and master's degrees at the Institute for Quantum Computing in Waterloo, Canada. 
	I observed a number of military contracts at my institution, which felt wrong to me. 
	Later, I became a Quaker---a pacifist religion---and became involved in anti-arms-trade activism. This is to say that I am personally committed against the proliferation of arms and the activities of imperial militaries. 

	Opposition to imperial war is a part of the history and culture of science. I am taking a position in this tradition, following the work of people like physicist Charles Schwartz who asked the American Physical Society (APS) to officially oppose the Vietnam War \citep{Schechner2000}. Hundreds more physicists joined this and related efforts under the banner of Science for the People \citep{Moore2008}. This work contributed to the creation of organizations today like the APS Physicists Coalition for Nuclear Threat Reduction \citep{AmericanPhysicalSociety2021}. 

	This history and my experience motivate me to conduct this analysis: I want more attention paid to military operations in scientific contexts to build more organized opposition to it. I am publishing this social science-informed work in a quantum-focused venue specifically to reach quantum workers who might resist warfare in QST.

\section{Theory: critical socio-political ethics \&\\ anti-imperialism}
	\label{sec:theory}


	Current research in ethics of QST focuses on approaches like responsible research and innovation (RRI), which attempt to ensure the social benefits of QST are maximized \citep[e.g.][]{TenHolter2021,Inglesant2018}. Some approaches aim to avoid QST which would worsen marginalization or cause harm broadly \citep[e.g.][]{Kop2021,Brennen2021}. This necessitates a social analysis of QST and a recognition of at least some of its potential harms \citep[e.g.][]{Moller2017,DeWolf2017}.

	However, these analyses generally miss the existence of power structures which are already causing immense harm, like imperialism. The reproduction of those power structures is concerning enough. This is a major critique of RRI approaches to QST, since RRI is typically conducted within capitalist frames. Some analyses have oriented themselves against some power structures but not others. For example, \cite{Brennen2021} note the dangers of ``reproduc[ing] and amplif[ying] existing socio-economic marginalisation'' (p. 11), yet in the same paper encourage more war and espionage involvement in the field (p. 24).

	I work instead with a critical socio-political approach to ethics \citep{Sloane2019,McLean2017}. This begins from the recognition that immense power structures shape technology and society, and from a desire to oppose these structures of oppression within and through the social and political nature of technology. This also changes the definition of ethics. The studies cited above typically deal with ethics of technology as identifying both the social influence of technology and ways to choose the best social influence. I build on Weston's approach to ethics, which takes as the central concern whether or not everyone needs are met \citep{Weston2018}. Needs and technologies are socially created in our politically unjust world, necessitating socio-political analysis \citep{Wajcman1991}. A critical socio-political ethics would seek to disrupt these pwoer structures---to not reproduce them, rather than to simply not make them worse. Crucially, this means that some QST may be better to not build at all.

	My analysis of military rhetoric is grounded by several points. First, I begin from anti-imperialism. The term imperialism describes political and cultural setups where one territory exerts ruling force over another; modern imperialism is often enacted through coups backed by foreign powers and corporations who extract materials and profits for the benefit of Western countries \citep{Said1994,Harvey2003}. America in particular maintains its power in large part through violent military interventions for the purpose of controlling resource extraction \citep{Butler1935,Mitchell2009,Christiansen2020,Forde2014,Harvey2003}. Anti-imperialism opposes all aspects of imperialism, including the immense, resource-intensive, technologically advanced American military. 

	Second, the linguistic foundation of my analysis is Cohn's \citeyearpar{Cohn1987} technostrategic language. This term describes how discussions in military scenarios typically have weapons and technology as the subjects, rather than people. This shifts questions about whether we ought to kill people into questions about how to design the best weapons for the sake of weapons (including sensing and communications technologies).

	Third, I contextualize the analysis of the AFRL in the growing body of military and nationalist think tank literature supporting the development of quantum technology for overtly imperialist purposes \citep[e.g.][]{Johnson2019,Kline2019,Smith2020a,Krelina2021,Kott2018,Herman2018,Guy2018}. The interest in QST is growing in military and intelligence circles. This work is often explicitly aimed at increasing the QST capability of America's military because China is increasing their QST capability, encouraging arms race activity \citep[e.g.][]{Kania2018,Kania2018a,Gamberini2021}. All of these cited works encourage the use of QST for war, and further encourage the glamour of the idea of types of quantum war. This is the background in which the AFRL provides the comments analyzed below.

\section{Site description \& methods} 
	\label{sec:site-methods}


	Quantum 2 Business (Q2B) is an annual conference run by the company QC Ware. I attended in December 2020, when the conference was hosted online for the first time. 
	The stated intent of the conference is to connect businesses to QST. The target audiences are business people who might use QST and people designing QST for businesses. It is a high-profile conference in the field, being sponsored and attended by all of the major QST corporations in North America: Google, IBM, Rigetti, Honeywell, and more.
	

	This is a site of significant military involvement in QST. The AFRL is highly involved---they were the main sponsor of the conference from 2018 through 2020, and have had multiple presentation slots every year. The opening remarks were delivered by Eric Schmidt, the former chairman of Google who has been providing advice to the US Department of Defense on technology adoption and providing funding to military startups for many years \citep{Conger2020}. 

	To explore the rhetoric of the AFRL in regards to quantum technology at this conference, I collected data by first watching all videos of the conference and taking detailed notes about content, visuals, and my personal reactions to talks. I took transcripts of all AFRL videos. I conducted a critical thematic analysis on the transcripts and written materials provided by the AFRL \citep{Lawless2019}. This analytic technique consists of reading data closely and noting pieces of language, or codes, that relate to ideas in a specific critical lens. I used the lenses of anti-imperialism (i.e. American military hegemony causes immense harm in the world and should be opposed) and technostrategic language (i.e. speaking in the language of technology/weapons structurally limits the ways that human suffering and death can be considered when talking about military actions). I collected codes while reading through each transcript several times. I organized these codes into groups of similar meaning or theme. Finally, I refined the groupings by comparing them to ideas in the grounding literature.

\section{Themes in the rhetoric of the AFRL}
	\label{sec:results}

	At Q2B 2020, the AFRL gave three talks \citep{V63,V64,V65}---referred to below as \emph{Talk 1}, \emph{Talk 2}, and \emph{Talk 3} respectively---and shared a pre-made video \citep{AFResearchLab2018}---\emph{Video 1}---and 3 documents on their sponsor page---the one referred to in this analysis is \emph{Document 1}. Italicized terms are used to make the distinction between data and literature clear.

	In these materials, the AFRL covered education programs they are sponsoring, start-up programs they are running, and the technology the AFRL is most interested in. This technology includes quantum cryptography, quantum clocks and other tech for GPS-free navigation, magnetometers for radar-free detection of submarines, and quantum computers and ML algorithms for operations decisions (i.e. tactical warfare decisions).

	Throughout these materials, four themes were clear.



	\subsubsection*{Theme 1: Quantum technology is for keeping the fight unfair}

		The mission statement of the AFRL, which was shared at Q2B, is to develop military technology that will ``keep the fight unfair'' (\emph{Talk 1, Slide 20})
		---that is, to maintain technological superiority of the American Air Force over other countries so they cannot fight back. QST is desired for maintaining ``national security and economic competitiveness'' (\emph{Talk 1, Slide 15}) 
		 and to ``better the nation at large'' (\emph{Talk 2}, 2:30). 

		In an imperial analysis of America, this means that the AFRL wants to use QST to uphold military and economic supremacy. According to the AFRL and in keeping with the techno-militarist logic of war profiteering, this goal requires constant upgrading of arms \citep{Butler1935,Duncan2013}.

		The AFRL is interested in many technologies, all for war. 
		These include magnetometers which can sense `bunker' structures \& submarines from long distances(\emph{Video 1}). In response to the possibility of jamming GPS signals and a desire for longer stealth operations, the AFRL wants sensitive accelerometers and clocks which maintain accurate timing for weeks so that they can go long times without GPS (\emph{Talk 1}, 7:13).

		Social arrangements like conferences and educational opportunities are also positioned as being explicitly for the AFRL's warfighting goals. The AFRL runs conferences that they call ``tech accelerators'' where students pitch technology ideas to the AFRL in part for their use (\emph{Talk 1}, 4:49).
		They encourage startups, allied countries, and academics to participate in their innovation ``paradigm'' (\emph{Talk 3}, 6:29).
		They acknowledge that this research is too challenging for the AFRL to take on alone, especially for fast deployment (\emph{Talk 1}, 17:08).
		Educational opportunities are similarly a ``springboard for collaboration'' for the production of war technology (\emph{Talk 2}, 0:30).

	\subsubsection*{Theme 2: Technostrategic language}

		Other than the motivating factors of ``national security and economic competitiveness'' (\emph{Talk 1, Slide 15}),
		the motivation to build particular technologies is cast as coming from physical properties. Passive language such as ``quantum sensors... will allow [some functionality]'' consistently positions the AFRL as simply a mediator in an inevitable technological outcome (\emph{Talk 1}, 7:16).
		One piece of written literature is entirely about how quantum properties directly translate to new technological applications, assuming from the start that the role of AFRL is to exploit quantum phenomena for new arms (\emph{Document 1}).

		This language focuses the discussion on the technology, and especially the technology-as-arms. The biggest questions are about, for example, how to make quantum sensors with the lowest size, weight, and power (SWaP) as possible (\emph{Talk 1, Slides 4, 6, 12}).
		Any questions about why are directed back to theme one: to give technology ``to [American] warfighters to meet their needs'' (\emph{Talk 2}, 1:01)
		to support the American military project.

		As \cite{Cohn1987} writes, techno-strategic language puts the speaker constantly in the seat of using weapons---not of being victimized by them. The subject is the quantum magnetometer, not the people in the building this sensor detects who are blown up. The subject is the quantum computer making rapid decisions about whether to strike with a drone, not the victims of the drone strike. Like Cohn, I found it harder to keep thinking about death when I spent enough time thinking about the technology---the language shifts our attention towards devices and away from humans. This language does not have words for horror.

	\subsubsection*{Theme 3: Revolution}

		AFRL representatives speak frequently about ``transformational technologies'' (\emph{Talk 2}, 1:01),
		disruption, revolution, and ``game-changing advances'' (\emph{Document 1}).
		The applications of quantum information are asserted to be ``life-altering'' (\emph{Talk 1}, 2:06).
		The AFRL's technologies have ``been transforming the Air Force [and] the world ... for many decades'' (\emph{Talk 1}, 1:06).
		Within this rhetoric is at least one contradiction: the internet (and the AFRL's role in its development as ARPAnet) is pointed to as an example of a technology whose impact could never have been predicted. It is used as an analogy to QST: QST will have an unpredictable social impact, which the AFRL predicts will be incomprehensibly large (\emph{Talk 1}, 16:36).

		The term ``second quantum revolution,'' coined by \cite{Dowling2003}, is used and discussed in quantum literature \citep{Roberson2021,Deutsch2020,Fox2020}. \cite{Roberson2021} argue that revolution and related terms have themselves military origins and ``frame the progress of science within the context of a `race' or `journey' frame. They are positive metaphors which imply advancement'' (391). There are ``ideals of technological progress that ensure economic and societal benefits and some inference of urgency'' \citep[p. 393]{Roberson2021}. This narrative does work for the AFRL by asserting that they will be providing a sure and good future for the nation against some urgent threat.


	\subsubsection*{Theme 4: Aesthetics of war}

		In its aesthetic presentation, the AFRL links QST to the practice of war. Commander Heather Pringle gave her presentation to Q2B in army briefs (\emph{Talk 3}).
		Each AFRL presentation began with the AFRL logo swooping in to dramatic music akin to what might be played at a tense moment in a first-person shooter video game (\emph{Talks 1, 2, \& 3}, 0:02).
		The term ``Q-Day'' is used by the AFRL to refer to the presumed time in the future when quantum technology will be deployed in the AF (\emph{Talk 1}).
		This references the military language of D-Day and other important dates identified by a letter \citep{HeadquartersDepartmentOfTheArmy2007}.

\begin{figure}
	     \centering
	     \begin{subfigure}[c]{0.45\textwidth}
	         \centering
	         \includegraphics[width=\textwidth]{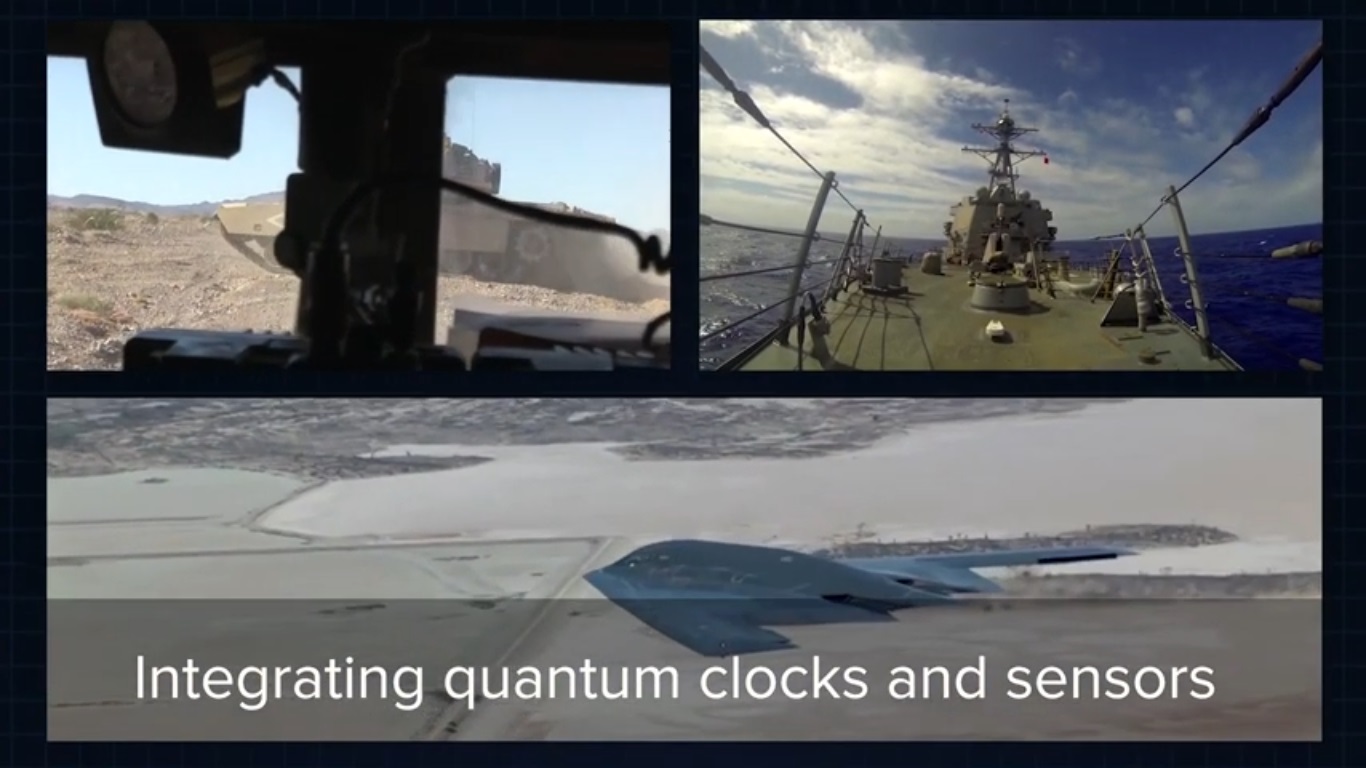}
	         \caption{}
	         \label{fig:quantum-clocks-war}
	     \end{subfigure}
	     \hfill
	     \begin{subfigure}[c]{0.45\textwidth}
	         \centering
	         \includegraphics[width=\textwidth]{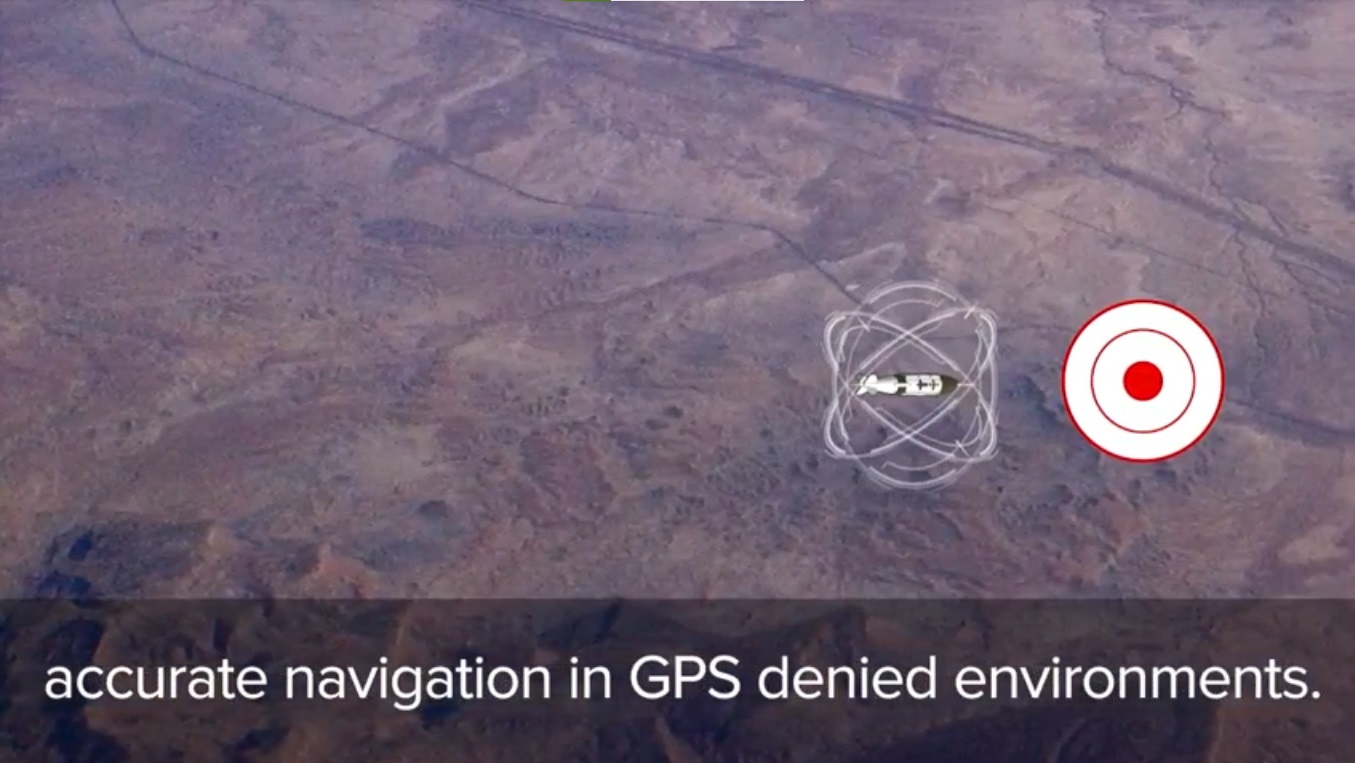}
	         \caption{}
	         \label{fig:target}
	     \end{subfigure}
	        \caption{Screenshots from the AFRL video `Quantum Information Science' \citep{AFResearchLab2018}. a) A view from an army vehicle in a bright rocky landscape, from the bow of a Navy ship, and a stealth aircraft from above. These are instances where the AFRL asserts quantum clocks and sensors can be used. b) A missile with cartoon electron orbitals around it is about to land on a target in a desert landscape.}
	        \label{fig:screenshots}
		\end{figure}

		The most striking example of the war aesthetic is in 15 seconds of a video titled `Quantum Information Science', published outside the scope of the Q2B conference but shared during it (\emph{Video 1}, 1:00--1:15).
		After describing quantum information in vague terms, rapid shots are shown of tanks, stealth aircraft, and aircraft carriers (see Fig. \ref{fig:quantum-clocks-war}). Subtitles describe how quantum technology will be able to assist in navigation in areas without GPS while a missile strikes a large red-and-white target in a desert landscape, all shown from an aerial view (see Fig. \ref{fig:target}). The desert landscape implies that the target is in the Middle East, with brown people the acceptable victims of war. The technology will also be able to assist in identifying bunkers, the subtitles go on, while a grainy image is shown of what might be a building covered in earth, superimposed with a target, and then blown up. This imagery viscerally ties specific quantum technologies to acts of war.

\section{Should QST support war?}
	\label{sec:discussion}


	In the previous section, I demonstrated that well-known elements of imperialist warfaring rhetoric are at play specifically in QST at the AFRL. That rhetoric is about the AFRL positioning themselves as a force for revolutionary imperial warfighting and capitalist gain, which they will use QST to enact. They associate QST with the aesthetics of the Air Force, which normalize and glamourize war. And they frame these `advances' as if they are driven by the technology itself, leaving out completely discussion of victims of war.

	I opened this paper by asking how the military is involved in QST. From what the AFRL has been discussing, it is clear at least that the military \emph{is} involved, and is trying to become more deeply involved. That means that QST could materially support imperial warfare in the near future.

	Do we as quantum scientists, engineers, and policy-makers want this science to prop up American imperialism? If we do not, then we should care about the path that QST is currently on. That path is one towards materially supporting devastating war.	Do we want QST and war to go hand-in-hand? If we do not, we must refuse images of QST being used for war, speak about the victims of empire, and support those already resisting empire.

	Do we want a QST which ignores critical social \& political analysis? If we do not, we must actively take up work against QST being used to maintain existing devastating power structures.

\section{Acknowledgements}

	Thanks to Juliane Borchert, Allison Gonsalves, Tara Roberson, and Elizabeth Patitsas for helpful comments.

\bibliographystyle{apalike}
\def\bibfont{\footnotesize}
\bibliography{Works-Cited}

\begin{thebibliography}{}

\bibitem[{AFResearchLab}, 2018]{AFResearchLab2018}
{AFResearchLab} (2018).
\newblock Quantum {Information} {Science}.

\bibitem[{AFRL}, 2020]{AF3}
{AFRL} (2020).
\newblock Quantum {Information} {Science}: {Precision} timing, sensing, secure
  networking, and exponentially faster processing.

\bibitem[{American Physical Society}, 2021]{AmericanPhysicalSociety2021}
{American Physical Society} (2021).
\newblock Physicists {Coalition} for {Nuclear} {Threat} {Reduction}.

\bibitem[{Australian Army}, 2021]{AustralianArmy2021}
{Australian Army} (2021).
\newblock Army {Quantum} {Technology} {Roadmap}.
\newblock Technical report.

\bibitem[Belcher et~al., 2020]{Belcher2020}
Belcher, O., Bigger, P., Neimark, B., and Kennelly, C. (2020).
\newblock Hidden carbon costs of the “everywhere war”: {Logistics},
  geopolitical ecology, and the carbon boot-print of the {US} military.
\newblock {\em Transactions of the Institute of British Geographers},
  45(1):65--80.
\newblock \_eprint:
  https://rgs-ibg.onlinelibrary.wiley.com/doi/pdf/10.1111/tran.12319.

\bibitem[Benjamin and Davies, 2018]{Benjamin2018a}
Benjamin, M. and Davies, N. J.~S. (2018).
\newblock The staggering death toll in {Iraq}.
\newblock {\em Salon}.
\newblock Section: News \& Politics.

\bibitem[Brennen et~al., 2021]{Brennen2021}
Brennen, G., Devitt, S., Roberson, T., and Rohde, P. (2021).
\newblock An {Australian} strategy for the quantum revolution.
\newblock page~36.

\bibitem[Butler, 1935]{Butler1935}
Butler, G.~S. (1935).
\newblock {\em War is a {Racket}}.
\newblock Round Table Press.

\bibitem[Christiansen, 2020]{Christiansen2020}
Christiansen, I. (2020).
\newblock Linkages {Between} {Economic} and {Military} {Imperialism}.
\newblock {\em World Review of Political Economy}, 11(3).

\bibitem[Cohn, 1987]{Cohn1987}
Cohn, C. (1987).
\newblock Sex and {Death} in the {Rational} {World} of {Defense}
  {Intellectuals}.
\newblock {\em Signs}, 12(4):687--718.
\newblock Publisher: University of Chicago Press.

\bibitem[Conger and Metz, 2020]{Conger2020}
Conger, K. and Metz, C. (2020).
\newblock ‘{I} {Could} {Solve} {Most} of {Your} {Problems}’: {Eric}
  {Schmidt}’s {Pentagon} {Offensive}.
\newblock {\em The New York Times}.

\bibitem[de~Wolf, 2017]{DeWolf2017}
de~Wolf, R. (2017).
\newblock The {Potential} {Impact} of {Quantum} {Computers} on {Society}.
\newblock {\em arXiv:1712.05380 [quant-ph]}.
\newblock arXiv: 1712.05380.

\bibitem[Deutsch, 2020]{Deutsch2020}
Deutsch, I.~H. (2020).
\newblock Harnessing the {Power} of the {Second} {Quantum} {Revolution}.
\newblock {\em PRX Quantum}, 1(2):020101.

\bibitem[Douglas, 2018]{Douglas2018}
Douglas, H.~E. (2018).
\newblock The {Bitter} {Aftertaste} of {Technical} {Sweetness}.
\newblock {\em Frankenbook}, page~6.

\bibitem[Dowling and Milburn, 2003]{Dowling2003}
Dowling, J.~P. and Milburn, G.~J. (2003).
\newblock Quantum technology: the second quantum revolution.
\newblock {\em Philosophical Transactions of the Royal Society of London.
  Series A: Mathematical, Physical and Engineering Sciences},
  361(1809):1655--1674.

\bibitem[Duncan and Coyne, 2013]{Duncan2013}
Duncan, T.~. and Coyne, C.~J. (2013).
\newblock The {Origins} of the {Permanent} {War} {Economy}.
\newblock {\em The Independent Review}, 18(2):219--240.

\bibitem[Forde, 2014]{Forde2014}
Forde, K. (2014).
\newblock Banana {Republic} {Blues}: {Militarization} and {Resources} in
  {Honduras}.
\newblock {\em NACLA Report on the Americas}, 47(1):10--13.

\bibitem[Fox et~al., 2020]{Fox2020}
Fox, M.~F., Zwickl, B.~M., and Lewandowski, H. (2020).
\newblock Preparing for the quantum revolution: {What} is the role of higher
  education?
\newblock {\em Physical Review Physics Education Research}, 16(2):020131.

\bibitem[Gamberini and Rubin, 2021]{Gamberini2021}
Gamberini, S.~J. and Rubin, L. (2021).
\newblock Quantum {Sensing}'s {Potential} {Impacts} on {Strategic} {Deterrence}
  and {Modern} {Warfare}.
\newblock {\em Orbis}, 65(2):354--368.

\bibitem[Guy, 2018]{Guy2018}
Guy, D.~C. (2018).
\newblock The weaponization of quantum physics: {How} quantum technology is
  transforming warfare.
\newblock Technical report.

\bibitem[Haraway, 1988]{Haraway1988}
Haraway, D. (1988).
\newblock Situated {Knowledges}: the science question in feminism and the
  privilege of partial perspective.
\newblock {\em Feminist Studies}, 14(3):575--599.

\bibitem[Harvey, 2003]{Harvey2003}
Harvey, D. (2003).
\newblock {\em The new imperialism}.
\newblock Oxford University Press, Oxford ; New York.
\newblock OCLC: ocm52784594.

\bibitem[Hayduk, 2020]{AF5}
Hayduk, M. (2020).
\newblock Quantum {Information} {Science} at {AFRL} - {Presentation} at {Q2B}.

\bibitem[{Headquarters, Department of the Army} and {United States Marine
  Corps}, 2007]{HeadquartersDepartmentOfTheArmy2007}
{Headquarters, Department of the Army} and {United States Marine Corps} (2007).
\newblock {FM} 101-5-1, {MCRP} 5-{2A} {Operational} {Terms} and {Graphics}.

\bibitem[Herman and Friedson, 2018]{Herman2018}
Herman, A. and Friedson, I. (2018).
\newblock Quantum {Computing}: {How} to {Address} the {National} {Security}
  {Risk}.
\newblock page~32.

\bibitem[Immerwahr, 2019]{Immerwahr2019}
Immerwahr, D. (2019).
\newblock {\em How to hide an empire: a history of the greater {United}
  {States}}.
\newblock Farrar, Straus and Giroux, New York, first edition edition.

\bibitem[Inglesant et~al., 2018]{Inglesant2018}
Inglesant, P., Jirotka, M., and Hartswood, M. (2018).
\newblock Responsible innovation in quantum technologies applied to defence and
  national security.
\newblock Technical report, Networked Quantum Information Technologies.

\bibitem[Johnson, 2019]{Johnson2019}
Johnson, W.~G. (2019).
\newblock {GOVERNANCE} {TOOLS} {FOR} {THE} {SECOND} {QUANTUM} {REVOLUTION}.
\newblock 59:35.

\bibitem[Kania, 2018]{Kania2018}
Kania, E.~B. (2018).
\newblock China’s {Quantum} {Quandary}.
\newblock {\em Military Cyber Affairs}, 3:13.

\bibitem[Kania and Costello, 2018]{Kania2018a}
Kania, E.~B. and Costello, J.~K. (2018).
\newblock Quantum {Hegemony}?: {China}'s ambitions and the challenge to
  {U}.{S}. innovation leadership.
\newblock page~52.

\bibitem[Kline et~al., 2019]{Kline2019}
Kline, K., Salvo, M., and Johnson, D. (2019).
\newblock How {Artificial} {Intelligence} and {Quantum} {Computing} are
  {Evolving} {Cyber} {Warfare}.

\bibitem[Kop, 2021]{Kop2021}
Kop, M. (2021).
\newblock Establishing a {Legal}-{Ethical} {Framework} for {Quantum}
  {Technology}.
\newblock page~16.

\bibitem[Kott et~al., 2018]{Kott2018}
Kott, A., Choi, K., Forch, B., Franaszczuk, P., Karna, S., Lee, S., Mait, J.,
  Reynolds, P., Sadler, B., Swami, A., and West, B. (2018).
\newblock Potential {Science} and {Technology} {Game} {Changers} for the
  {Ground} {Warfare} of 2050: {Selected} {Projections} {Made} in 2017.
\newblock page~42.

\bibitem[Krelina, 2021]{Krelina2021}
Krelina, M. (2021).
\newblock Quantum {Warfare}: {Definitions}, {Overview} and {Challenges}.
\newblock {\em arXiv:2103.12548 [physics, physics:quant-ph]}.
\newblock arXiv: 2103.12548.

\bibitem[Kumar, 2017]{Kumar2017}
Kumar, D. (2017).
\newblock National {Security} {Culture}: {Gender}, {Race}, and {Class} in the
  {Production} of {Imperial} {Citizenship}.
\newblock {\em International Journal of Communication}, 11:2154--2177.

\bibitem[Lawless and Chen, 2019]{Lawless2019}
Lawless, B. and Chen, Y.-W. (2019).
\newblock Developing a {Method} of {Critical} {Thematic} {Analysis} for
  {Qualitative} {Communication} {Inquiry}.
\newblock {\em Howard Journal of Communications}, 30(1):92--106.

\bibitem[McLean et~al., 2017]{McLean2017}
McLean, S.-A., Thomas, N., and Okakpu, O.~K. (2017).
\newblock D.{A}.{T}.{S} {Scientific} {Ethics} {Statement} \& {Reading} {Guide}.
\newblock {\em Decolonize All The Science}.

\bibitem[Mitchell, 2009]{Mitchell2009}
Mitchell, T. (2009).
\newblock Carbon democracy.
\newblock {\em Economy and Society}, 38(3):399--432.

\bibitem[Möller and Vuik, 2017]{Moller2017}
Möller, M. and Vuik, C. (2017).
\newblock On the impact of quantum computing technology on future developments
  in high-performance scientific computing.
\newblock {\em Ethics and Information Technology}, 19(4):253--269.

\bibitem[Mohanty et~al., 2008]{Mohanty2008}
Mohanty, C.~T., Riley, R.~L., and Pratt, M.~B., editors (2008).
\newblock {\em Feminism and war: confronting {US} imperialism}.
\newblock Zed Books, London ; New York.
\newblock OCLC: ocn191925652.

\bibitem[Moore, 2008]{Moore2008}
Moore, K. (2008).
\newblock {\em Disrupting science: social movements, {American} scientists, and
  the politics of the military, 1945-1975}.
\newblock Princeton studies in cultural sociology. Princeton University Press,
  Princeton.
\newblock OCLC: ocn128237783.

\bibitem[{National Defence Canada}, 2021]{NationalDefenceCanada2021}
{National Defence Canada} (2021).
\newblock {DND}/{CAF}’s {Quantum} {Science} and {Technology} {Strategy}.
\newblock Last Modified: 2021-01-18.

\bibitem[{Office of Naval Research}, 2021]{OfficeOfNavalResearch2021}
{Office of Naval Research} (2021).
\newblock Programs - {Quantum} {Information} {Science} - {Office} of {Naval}
  {Research}.

\bibitem[{QC Ware}, 2021a]{V65}
{QC Ware} (2021a).
\newblock {Q2B} 2020 {\textbar} {AFRL} {Overview} and {AFWERX} {Summary}
  {\textbar} {Brigadier} {General} {Heather} {L}. {Pringle} and more.

\bibitem[{QC Ware}, 2021b]{V64}
{QC Ware} (2021b).
\newblock {Q2B} 2020 {\textbar} {Quantum} {Education} {\textbar} {Tim}
  {Lawrence} {\textbar} {Air} {Force} {Research} {Laboratory}.

\bibitem[{QC Ware}, 2021c]{V63}
{QC Ware} (2021c).
\newblock {Q2B} 2020 {\textbar} {Quantum} {Information} {Science} at the {Air}
  {Force} {Research} {Laboratory} {\textbar} {Michael} {Hayduk} {\textbar}
  {AFRL}.

\bibitem[Roberson et~al., 2021]{Roberson2021}
Roberson, T., Leach, J., and Raman, S. (2021).
\newblock Talking about public good for the second quantum revolution:
  analysing quantum technology narratives in the context of national
  strategies.
\newblock {\em Quantum Science and Technology}, 6(2):025001.

\bibitem[Said, 1994]{Said1994}
Said, E.~W. (1994).
\newblock {\em Culture and imperialism}.
\newblock Vintage Books, New York, 1st vintage books ed edition.

\bibitem[Schechner et~al., 2000]{Schechner2000}
Schechner, S., Ripin, B., Lustig, H., and Wilson, R.~M. (2000).
\newblock Consciousness {Raising}.
\newblock {\em APS News}, 9(1):2.

\bibitem[Schuster, 2015]{Schuster2015}
Schuster, E.~A. (2015).
\newblock Transistorized {State}: {The} {Technique} of {Hegemony} in the
  {Early} {Cold} {War}.
\newblock {\em Perspectives on Global Development and Technology},
  14(1-2):178--189.

\bibitem[Sloane, 2019]{Sloane2019}
Sloane, M. (2019).
\newblock Inequality {Is} the {Name} of the {Game}: {Thoughts} on the
  {Emerging} {Field} of {Technology}, {Ethics} and {Social} {Justice}.
\newblock {\em Weizenbaum Conference}.
\newblock Publisher: WI - Weizenbaum Institute for the Networked Society
  Version Number: 1.

\bibitem[Smith, 2020]{Smith2020a}
Smith, M. S. D.~M. (2020).
\newblock Warfare: 2050.
\newblock {\em NCO Journal}, pages 1--5.

\bibitem[Ten~Holter et~al., 2021]{TenHolter2021}
Ten~Holter, C., Inglesant, P., and Jirotka, M. (2021).
\newblock Reading the road: challenges and opportunities on the path to
  responsible innovation in quantum computing.
\newblock {\em Technology Analysis \& Strategic Management}, pages 1--13.

\bibitem[Van~Keuren, 1997]{VanKeuren1997}
Van~Keuren, D.~K. (1997).
\newblock Science {Goes} to {War}: {The} {Radiation} {Laboratory}, {Radar}, and
  {Their} {Technological} {Consequences}.
\newblock {\em Reviews in American History}, 25(4):643--647.
\newblock Publisher: Johns Hopkins University Press.

\bibitem[Wajcman, 1991]{Wajcman1991}
Wajcman, J. (1991).
\newblock {\em Feminism {Confronts} {Technology}}.
\newblock Pennsylvania State University Press.

\bibitem[Weston, 2018]{Weston2018}
Weston, A. (2018).
\newblock {\em A 21st century ethical toolbox}.
\newblock Oxford University Press, New York, fourth edition edition.

\end{thebibliography}

\end{document}